\newif\ifblind
\blindtrue          

\documentclass[conference]{IEEEtran}

\usepackage[T1]{fontenc}
\usepackage[utf8]{inputenc}
\usepackage{times}            
\usepackage{microtype}        

\usepackage{amsmath}
\usepackage{amssymb}
\usepackage{amsthm}

\usepackage{graphicx}
\usepackage{subcaption}       
\usepackage{epsfig}
\usepackage{svg}

\usepackage{booktabs}         
\usepackage{multirow}
\usepackage{tabularx}

\usepackage{algorithm}
\usepackage{algpseudocode}
\usepackage[hidelinks]{hyperref}

\usepackage{listings}
\usepackage{xcolor}
\usepackage{booktabs,multirow}

\usepackage{hyperref}
\hypersetup{
    pdfauthor={Anonymous},      
    pdfsubject={IEEE S\&P 2026},
    pdfkeywords={security, privacy}
}
\usepackage{url}

\usepackage{balance}          
\usepackage{enumitem}         
\usepackage{xspace}           
\usepackage{cleveref}         

\theoremstyle{plain}

\theoremstyle{definition}

\theoremstyle{remark}

\begin{document}

\title{%
    FLINT: Fingerprinting Federated Learning Architectures from 5G PHY-Layer Side Channels  
}



\author{\IEEEauthorblockN{Md Nahid Hasan Shuvo, Mahmudul Hassan Ashik, Moinul Hossain} 
\IEEEauthorblockA{George Mason University, Fairfax, VA, USA}\IEEEauthorblockA{ Email: mshuvo@gmu.edu, mashik@gmu.edu, mhossa5@gmu.edu}}

\maketitle

\begin{abstract}

Federated Learning (FL) over 5G cellular networks protects raw data but remains vulnerable to side-channel leakage. Prior fingerprinting attacks assume packet-level network visibility, an assumption that does not hold at the 5G Physical (PHY) layer, where user payloads are encrypted and Radio Network Temporary Identifiers (RNTIs) may change over time. However, we demonstrate that PHY-layer scheduling metadata broadcast over the Physical Downlink Control Channel (PDCCH) preserves architecture-associated temporal patterns. We introduce FLINT, a novel black-box fingerprinting framework that infers FL model architecture families, including CNNs, RNNs, and Transformers, using only coarse PHY-layer observations. FLINT overcomes the lack of network-layer visibility by decoding PDCCH scheduling information, mapping changing RNTIs to physical user devices, and applying multi-view temporal modeling to distinguish architecture-specific training behavior. This leakage is security-critical because knowledge of a client’s model architecture can transform passive reconnaissance into targeted downstream exploitation. Extensive experiments on an over-the-air srsRAN-based 5G testbed demonstrate that FLINT achieves a macro F1-score of 0.930 for architecture-family classification. To our knowledge, FLINT is the first work to fingerprint AI/ML model architectures using lower- layer 5G side-channel information obtainable by any protocol-aware adversary.

\end{abstract}


\section{Introduction}

Federated learning (FL) has emerged as a widely adopted paradigm for training deep learning models across distributed devices without collecting their raw data at a central server~\cite{mcmahan2017fl,kairouz2021advances}. By keeping data on-device and exchanging only encrypted model updates, FL has been adopted in privacy-sensitive domains such as healthcare, mobile robotics, and the Internet of Things. Increasingly, these deployments rely on Fifth Generation (5G) cellular networks, which provide the low-latency and wide-area connectivity required by geographically distributed clients. Encryption, however, protects only the contents of the updates, not the communication patterns by which they are exchanged. A growing body of work has shown that metadata associated with encrypted traffic, including packet sizes, directions, and timing, leaks significantly more information than intended. Such information enables adversaries to fingerprint visited websites~\cite{wang2020wf}, identify mobile applications~\cite{li2024app}, and infer device types~\cite{chowdhury2022iot} through passive observation alone.

Recently, this insight has been extended to the federated learning setting. FLARE~\cite{flare} demonstrated that a passive adversary observing encrypted Wi-Fi traffic can infer the architecture of the model being trained, distinguishing convolutional neural networks (CNNs) from recurrent neural networks (RNNs). This capability has important security implications, as knowledge of a client's model architecture enables targeted downstream attacks, including architecture-specific adversarial examples, model inversion attacks, and resource denial attacks~\cite{zhou2024adv,ashik2025reaper}. However, FLARE and related studies assume access to network-layer packet information, including packet sizes, directions, and inter-arrival times. While this assumption holds in Wi-Fi networks, it does not hold in 5G.

In 5G networks, user-plane traffic is protected by the Access Stratum security context, preventing external observers from accessing network-layer packets. Instead, the only information visible over the air is physical-layer control signaling. In particular, scheduling metadata transmitted on the Physical Downlink Control Channel (PDCCH) remains observable. For each scheduling decision, the base station broadcasts Downlink Control Information (DCI) records specifying the scheduled device, allocated resource blocks, and corresponding transport block size. These records are transmitted without encryption and can be decoded by passive software-defined radio receivers without network credentials~\cite{wan2024nrscope,ludant20235g}. Whether this coarser signal still carries sufficient information to fingerprint a client's model architecture remains unexplored, and we address this question in this paper.

\textbf{Challenges:} Answering this question presents three key challenges. First (\textbf{C1}), a passive observer in 5G has access only to unlabeled physical-layer scheduling records, which must be mapped back to individual clients despite rotating identifiers and the complexity of operating in a real 5G testbed environment. This complicates the construction of consistent per-device activity streams under live network conditions. Second (\textbf{C2}), packet-level traffic information is not available at the physical layer. Instead, the observer sees only scheduler-generated signals such as resource allocations and transport block sizes. These measurements are indirect and influenced by channel conditions and scheduling decisions, making them a coarse proxy for communication activity. Third (\textbf{C3}), traces are inherently incomplete due to practical constraints in over-the-air collection and decoding, leading to missing or partially observed scheduling records. As a result, the adversary must infer structure from irregular temporal observations. Finally (\textbf{C4}), simple statistics based on communication activity are insufficient to distinguish between model families, as different architectures can induce similar overall transmission patterns. This requires representations that capture higher-order temporal dynamics.

Addressing these challenges requires a fingerprinting approach that operates on incomplete, coarse-grained, and temporally irregular physical-layer observations. Building on this insight, we develop \textsc{Flint}, a passive fingerprinting framework for identifying federated learning model architectures from 5G PHY-layer side-channel information.

\textbf{Contributions:}
We investigate whether federated learning model architectures, including CNNs, RNNs, and Transformers, can be inferred using only 5G physical-layer scheduling metadata, and show that capturing temporal structure is essential.

In summary, our contributions are as follows:

\begin{itemize}

\item We present a fully black-box data collection pipeline on a real srsRAN-based 5G testbed, including an algorithm that maps rotating physical-layer identifiers (RNTIs) back to individual client devices using only the timing of decoded records, addressing \textbf{C1}.

\item We identify a new physical-layer side channel in 5G federated learning, in which scheduling metadata broadcast over the PDCCH leaks information about the underlying learning process and can be exploited by an adversary without network-layer visibility, thereby addressing \textbf{C2}.

\item We design a fingerprinting framework that operates on incomplete and irregular scheduling observations and captures the temporal evolution of training behavior, improving robustness under realistic 5G conditions, addressing \textbf{C3}.

\item We show that simple communication statistics derived from resource block allocations and transport block sizes are insufficient to distinguish among model families, and we evaluate robustness under different abalation studies to address \textbf{C4}.

\end{itemize}

The remainder of this paper is organized as follows. Section~\ref{sec:related} discusses the current related research. Section~\ref{sec:background} provides background on 5G cell attachment and DCI decoding. Section~\ref{sec:threat_model} shows our threat model. Section~\ref{sec:methodology} presents the system model and our fingerprinting framework pipeline. Section~\ref{sec:evaluation} describes the dataset and results and shows analysis. Finally, Section~\ref{sec:conclusion} concludes.\vspace{-0.05in}

\section{Related Work}\label{sec:related}

This section reviews prior work on privacy attacks in federated learning, traffic fingerprinting, side-channel inference of model properties, and 5G physical-layer monitoring.

\subsection{Privacy Attacks on Federated Learning}

Federated learning was introduced to keep training data on-device~\cite{mcmahan2017fl}, but extensive prior work has shown that shared model updates can still leak sensitive information. Membership inference attacks determine whether a specific record was used during training~\cite{shokri2017membership,nasr2019comprehensive}, and have been extended to federated settings by exploiting per-round gradients exchanged during aggregation~\cite{zhu2025fedmia}. Gradient inversion attacks reconstruct representative training samples directly from shared updates~\cite{yang2025inversion,chen2025diffusion}, while property inference attacks recover aggregate attributes of local datasets~\cite{mehnaz2022sensitive,wang2022poisoning}.

These attacks typically assume access to model updates or participation in the learning process. In contrast, our work considers a fully passive adversary that neither observes gradients nor participates in training, and instead infers the model architecture from over-the-air observations.

\subsection{Traffic Fingerprinting and Side-Channel Leakage}

Traffic analysis has been widely used to infer sensitive information from encrypted communications. Website fingerprinting attacks identify visited pages over Tor or HTTPS using packet sizes, directions, and timing information~\cite{wang2020wf,sirinam2018df,panchenko2016wf}, and remain effective under various defenses~\cite{shen2023rf,deng2025countmamba}. Related approaches infer mobile application usage~\cite{li2024app} and identify IoT device types from network behavior~\cite{chowdhury2022iot,sheng2025iiot}.

These methods rely on network-layer visibility and access to per-packet information. FLARE~\cite{flare} applies similar ideas to federated learning using packet-level statistics of encrypted Wi-Fi traffic to distinguish neural architectures such as CNNs and RNNs. However, it relies on network-layer observability that is not available in 5G physical-layer settings. We further show that coarse communication statistics are insufficient to distinguish among modern model families based on physical-layer observations.

\subsection{Recovering Model Properties via Side Channels}

A related line of work studies the recovery of neural network properties from hardware and system-side channels. Electromagnetic emissions have been used to infer network structure and parameters~\cite{batina2019csinn}, while power-based side channels reveal architectural details in controlled environments~\cite{zhang2023power,chabanne2021side}. Timing-based attacks can also expose information about dataset size and model complexity~\cite{shokri2017membership}.

These approaches typically assume physical access, co-location, or controlled execution environments, and are applied to centralized training or inference on a single machine. In contrast, our work targets a distributed federated learning setting and infers architecture remotely using only physical-layer side-channel metadata.

\subsection{5G Physical-Layer Monitoring and PDCCH Sniffing}

The openness of the 5G control plane has enabled passive monitoring of scheduling behavior. Tools such as NR-Scope~\cite{wan2024nrscope} and 5GSniffer~\cite{ludant20235g} demonstrate that the Physical Downlink Control Channel (PDCCH) can be blind-decoded by unprivileged receivers, revealing per-device scheduling decisions, including resource allocations and transport block sizes.

Prior work has used this visibility to infer coarse user activity patterns, analyze information leakage in wireless systems, and enable control-channel-based attacks~\cite{ashik2025reaper}. However, no prior study has used decoded PDCCH metadata to infer federated learning model architectures. Our work builds on these decoding capabilities as the data-collection substrate and addresses the challenge of reconstructing per-device traces from interleaved scheduling records with rotating physical-layer identifiers.\vspace{-0.05in}


\section{Background}
\label{sec:background}\vspace{-0.05in}

\subsection{Adversarial View of Cell Attachment and DCI Decoding}
\label{subsec:initial_attach}\vspace{-0.05in}

A UE that powers \emph{ON} (or toggles airplane mode \emph{OFF}) must complete cell search, system-information acquisition, random access, and \emph{Radio Resource Control (RRC)} connection establishment before exchanging user data, and every one of these steps occurs in plaintext on channels that a passive receiver can decode without credentials. We therefore describe attachment from the adversary's vantage: it performs the same synchronization and decoding as a legitimate UE but stops short of completing random access or establishing a security context.

\noindent\textbf{Synchronization and system information.}
The receiver captures in-phase/quadrature (IQ) samples at the cell center frequency and aligns to the $10\,$ms frame by locating the \emph{Synchronization Signal Block (SSB)}, whose \emph{Primary} and \emph{Secondary Synchronization Signals (PSS/SSS)} yield timing, frequency offset, and the \emph{Physical-layer Cell Identity (PCI)}~\cite{3gpp-ts38211}. Decoding the \emph{Physical Broadcast Channel (PBCH)} returns the \emph{Master Information Block (MIB)}, which locates \emph{Control Resource Set~0 (CORESET\#0)} and the Type0-PDCCH search space. The receiver then decodes \emph{SIB1} on the \emph{Physical Downlink Shared Channel (PDSCH)}, scheduled by \emph{Downlink Control Information (DCI)} scrambled with the standardized \emph{System Information RNTI (SI-RNTI)} on the \emph{Physical Downlink Control Channel (PDCCH)}~\cite{3gpp-ts38331,3gpp-ts38213}. SIB1 exposes the random-access and common radio configuration (e.g., \emph{PRACH} parameters, \texttt{offsetToPointA}), which is sufficient to monitor the cell without authentication.

\begin{figure}[t] 
\centering
\includegraphics[width=0.9\columnwidth]{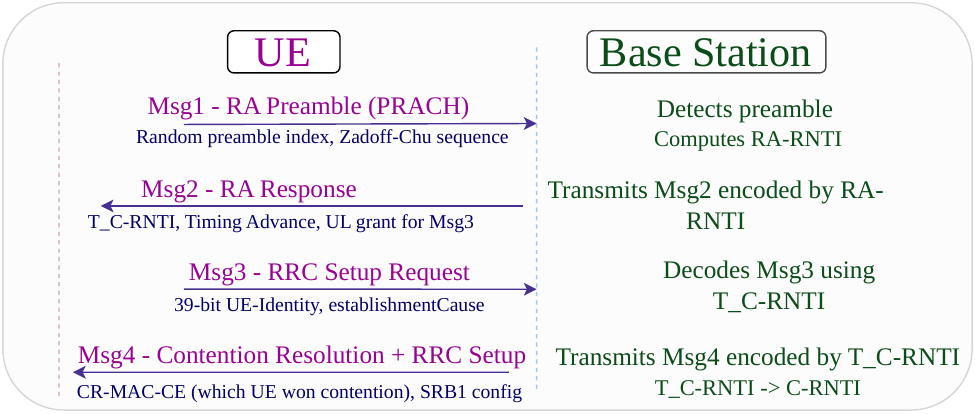} 
\caption{Random access establishment. Msg1--Msg4 and all RRC up to and including \emph{RRCSetup} and can be decoded by anyone with protocol knowledge}\vspace{-0.25in} 
\label{fig:rrcConnection}
\end{figure}

\noindent\textbf{Random access and the plaintext window.}
Fig.~\ref{fig:rrcConnection} shows the four-message random-access exchange and marks the adversary's observation window. The UE sends a \emph{PRACH} preamble (\textbf{Msg1}); the gNB replies with the \emph{Random Access Response} (\textbf{Msg2}, scrambled by \emph{RA-RNTI}) carrying a Timing Advance, a Msg3 grant, and a \emph{Temporary C-RNTI (TC-RNTI)}~\cite{3gpp-ts38321}. The UE then transmits \emph{\textbf{RRCSetupRequest}} (\textbf{Msg3}) and receives \emph{\textbf{RRCSetup}} (\textbf{Msg4}, scrambled by TC-RNTI); the \emph{Contention Resolution MAC-CE} echoes the winning UE's identity, namely a random value at first attach, an \emph{S-TMSI} on reconnection, or a \emph{C-RNTI} on handover~\cite{3gpp-ts38321}. The winner's TC-RNTI is then promoted to a \emph{C-RNTI} and the UE completes \emph{\textbf{RRCSetupComplete}}, establishing \emph{Signaling Radio Bearer~1 (SRB1)}. NAS authentication and the \emph{NAS} and \emph{AS Security Mode Command (SMC)} procedures follow~\cite{3gpp-ts33501}. Every RRC message up to and including Msg4 is sent in plaintext; the adversary's observation window closes once the \emph{AS-SMC} activates ciphering on the radio bearer. DCI scheduling metadata on the PDCCH, however, is \emph{never} AS-ciphered~\cite{3gpp-ts33501}, so per-grant scheduling records remain observable for the entire session.

\noindent\textbf{Blind decoding of the DCI stream.}
\label{subsec:dci_decoding}%
The gNB confines PDCCH to CORESETs whose locations are advertised in MIB and SIB1, so a receiver that has decoded both can reconstruct every candidate slot and \emph{blind-decode} it: it applies polar decoding to each candidate and checks the appended 24-bit CRC, which is scrambled with the recipient's RNTI, so a clean CRC simultaneously validates the message \emph{and} reveals the target RNTI~\cite{3gpp-ts38212}. Broadcast messages (SI-RNTI and paging \emph{P-RNTI}) decode without any per-device identifier; UE-specific grants require the C-RNTI range obtained during reconnaissance (\S\ref{subsec:data_collect}). The output is a time-ordered stream of DCI records, each a tuple of timestamp, RNTI, direction, and scheduling fields.

\paragraph{\textbf{Vulnerability}} Due to a lack of any security protocol at the start of the RACH process, any UE with a SIM (registered or not) can trigger the RACH procedure from the base station: the SIM is not authenticated until RRC establishment completes and the UE can reach the core network. An adversary can therefore force the base station to disclose critical configuration about itself (e.g., available TC-RNTI, CORESET\#1 parameters, DCI format) simply by triggering RACH, with no registered SIM required. We show in \S\ref{subsec:data_collect} how we exploit this to trigger RACH and obtain the parameters needed to decode UE-specific DCI.

\subsection{Federated Learning over 5G and Its Side-Channel Surface}
\label{subsec:fl-surface}

Having described how an adversary decodes DCI records from the PDCCH, we now explain why those records carry information about the model architecture being trained. The connection arises from how a federated training session is mapped to the radio scheduler.

\noindent\textbf{The federated round as a communication pattern.}
We consider synchronous federated learning, the predominant setting on mobile and edge deployments and the one used in our testbed~\cite{mcmahan2017fl}. Training proceeds in discrete rounds: the server broadcasts the current global model; each client trains locally for several epochs and returns an update; the server waits for all selected clients, aggregates, and broadcasts the next model. This yields a repeating per-client pattern, a downlink burst as the global model arrives, near-silence during local computation, and an uplink burst as the update is sent. Two quantities of this pattern are set largely by the client's model rather than the network. The \emph{size} of each update scales with the trainable-parameter count $\theta$ (a single update is on the order of $4\theta$ bytes for 32-bit floats), and the \emph{duration} of the silent interval scales with the cost of a forward/backward pass, which differs across architectures even at comparable $\theta$. Architecture therefore shapes both how much a client transmits per round and the rhythm with which it does so.

\noindent\textbf{Mapping the round onto 5G scheduling.}
At the physical layer, this pattern is observed not as packets but as scheduling grants. An upload appears as a series of uplink DCI grants whose transport-block sizes accumulate to the update volume; the global-model arrival appears symmetrically as downlink grants; the local-computation interval appears as a gap with few or no grants. Because the adversary decodes every grant addressed to the client (\S\ref{sec:methodology}), it observes a time-ordered stream of transport-block sizes and directions tracing the upload--compute--download structure of each round, without ever seeing a user-plane packet. 

\noindent\textbf{Why architecture leaks, and why volume is not enough.}
Architecture influences the signal through two channels. The first is \emph{volume}: larger parameter counts produce larger transport-block totals per round. The second is \emph{temporal structure}: the cadence of rounds, the regularity of the upload--compute--download cycle, and the way successive updates evolve. Volume alone is a strong cue when update sizes are distinctive, but it becomes ambiguous when two architecture families have similar parameter counts, as is typical for the compact models used at the edge. In that regime, the temporal structure, not the volume, separates the families, which motivates the feature design of \S\ref{sec:methodology}.
\vspace{-0.05in}

\section{Threat Model}
\label{sec:threat_model}

We consider an external adversary operating within the coverage area of a 5G cell in a federated learning deployment. The adversary does not participate in training, does not transmit any interference signal during the fingerprinting stage, and has no access to base station infrastructure or client devices. All user-plane communications, including model updates, are protected by the 5G security architecture and remain inaccessible to the adversary.

\subsection{Adversarial Goal}

The adversary aims to infer the underlying federated learning model architecture used by participating clients from over-the-air physical-layer observations.

\subsection{Adversarial Capabilities}

\smallskip\noindent\textbf{C1. Physical-layer observability.}
The adversary can passively observe and decode downlink control information (DCI) transmitted on the Physical Downlink Control Channel (PDCCH). These observations include scheduling decisions such as the scheduled user identity, allocated resource blocks, and transport block sizes.

\smallskip\noindent\textbf{C2. Passive monitoring.}
The adversary operates in a fully passive manner using standard software-defined radio reception capabilities. It does not inject any adversarial traffic, interfere with network operation, or modify any transmissions.

\subsection{Adversarial Limitations}

\smallskip\noindent\textbf{L1. No access to user-plane data.}
The adversary cannot observe or reconstruct network-layer packets, including packet payloads, sizes, directions, or timing information. All user data and model updates remain encrypted under the 5G security framework.

\smallskip\noindent\textbf{L2. Restricted observation model.}
The adversary is limited to physical-layer scheduling metadata obtained from decoded DCI messages. These observations are coarse-grained and do not expose packet-level structure or application-layer semantics.\vspace{-0.05in}

\section{Methodology}
\label{sec:methodology}

This section describes the end-to-end attack pipeline: from passive over-the-air reconnaissance and DCI decoding, through RNTI-to-device mapping via the Conflict-Constrained Chain Decomposition (CCCD) algorithm, to the construction of per-device feature datasets suitable for AI/ML fingerprinting.

\begin{figure*}[ht]
    \centering
    \includegraphics[width=\textwidth]{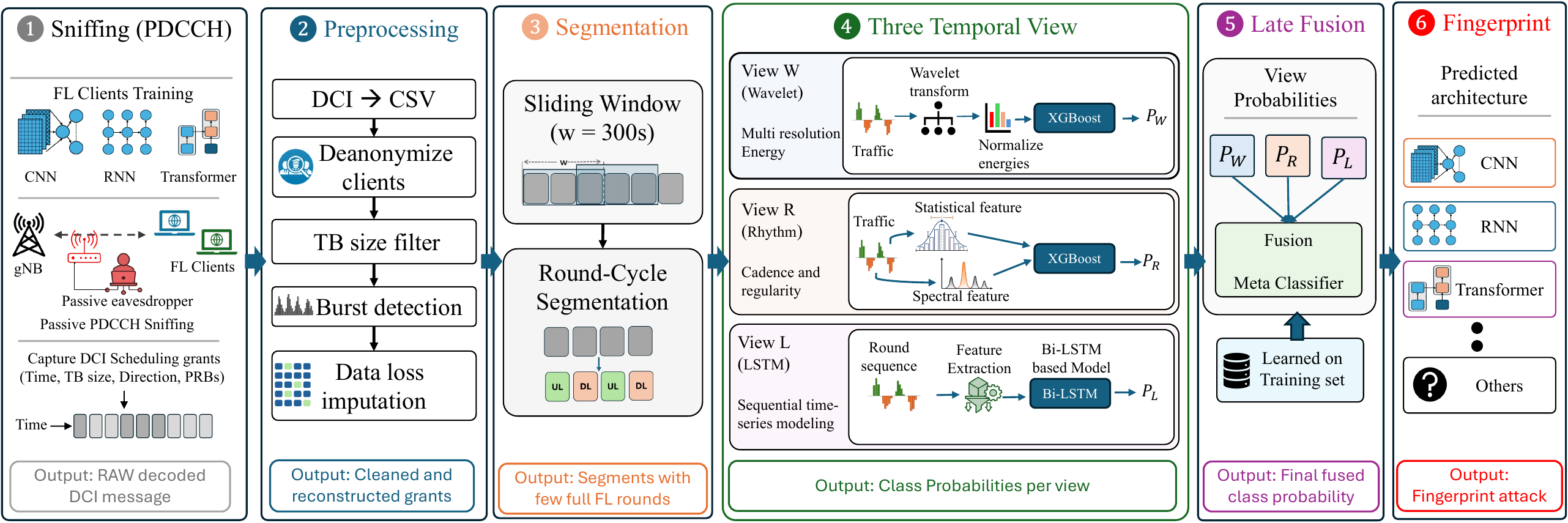}
    \caption{Proposed system pipeline, detailing the steps from PDCCH sniffing to fingerprinting.} \vspace{-0.20in}
    \label{fig:system}
\end{figure*}

\subsection{System Model and Attack Pipeline}
\label{subsec:system-model}

\noindent\textbf{Deployment model.}
We consider a 5G cell served by a single gNodeB (gNB), where three client devices participate in a federated learning session. Each client is a User Equipment (UE) device that trains a deep learning model locally and exchanges model updates with a federated server over the 5G network. The gNB schedules uplink and downlink transmissions through the Physical Downlink Control Channel (PDCCH), where scheduling decisions are carried as Downlink Control Information (DCI) records. Given decoded physical-layer scheduling records, the objective is to infer, for each observed client, the model architecture family being trained.

\noindent\textbf{Architecture families.}
We study three architecture families that commonly appear in federated learning workloads: convolutional neural networks (CNNs), recurrent neural networks (RNNs), and Transformer-based models. Each family contains multiple concrete model variants. For example, the CNN family may include both custom networks and standard backbones such as ResNet and DenseNet. Thus, the fingerprinting target is not a specific model instance, but the broader architecture family. In the open-world setting, we additionally include an \emph{Others} category for traffic that does not belong to the trained FL families, such as background non-FL activity or unseen workloads.

\noindent\textbf{Observation model.}
After DCI decoding and per-device separation, the adversary obtains a time-ordered sequence of scheduling records for each client. We denote a client's observation stream as
\[
S = {(t_i, d_i, b_i, p_i)}_{i=1}^{n},
\]
where $t_i$ denotes the timestamp of the $i$-th scheduling record,
$d_i \in {\text{UL}, \text{DL}}$ denotes the transmission direction,
$b_i$ denotes the transport block size in bytes, and $p_i$ denotes the
number of allocated physical resource blocks. The stream $S$ is segmented
into fixed-duration observation windows. For each window, the objective is to infer
\[
\hat{y} \in {\text{CNN}, \text{RNN}, \text{Transformer}, \text{Others}}.
\]
In the closed-world setting, the task is three-way classification among CNN, RNN, and Transformer when the window is known to belong to one of the trained FL families. In the open-world setting, the classifier must also reject windows that do not match any known FL family and assign them to \emph{Others}.

\noindent\textbf{Pipeline overview.}
Figure~\ref{fig:system} summarizes the end-to-end \textsc{Flint} pipeline. The pipeline begins with passive PDCCH sniffing, which produces raw decoded DCI records. The decoded records are converted into a structured table and preprocessed through transport block filtering, burst detection, data-loss imputation, and RNTI-to-device mapping to reconstruct client-specific traces. The cleaned per-device stream is then segmented into fixed-duration observation windows and further decomposed into round-cycle segments. Each window is represented through three complementary temporal views: a wavelet-based multi-resolution energy view (W), a rhythm view that captures cadence and regularity (R), and a Bi-LSTM-based sequential modeling view that captures round-to-round evolution (L). Each view produces a class-probability vector, and a late-fusion meta-classifier combines these view-level probabilities into the final architecture prediction.

Stages related to raw DCI collection, preprocessing, data-loss imputation, and RNTI-to-device mapping are described in Sections~\ref{subsec:data_collect} and~\ref{sec:cccdh_overview}. The remaining stages, including segmentation, temporal view extraction, late fusion, and fingerprint decisions, constitute the fingerprinting framework described next.
\vspace{-0.05in}

\subsection{Black-box Data Collection Pipeline}
\label{subsec:data_collect}

\begin{figure}[t]
  \centering
  \includegraphics[width=\columnwidth]{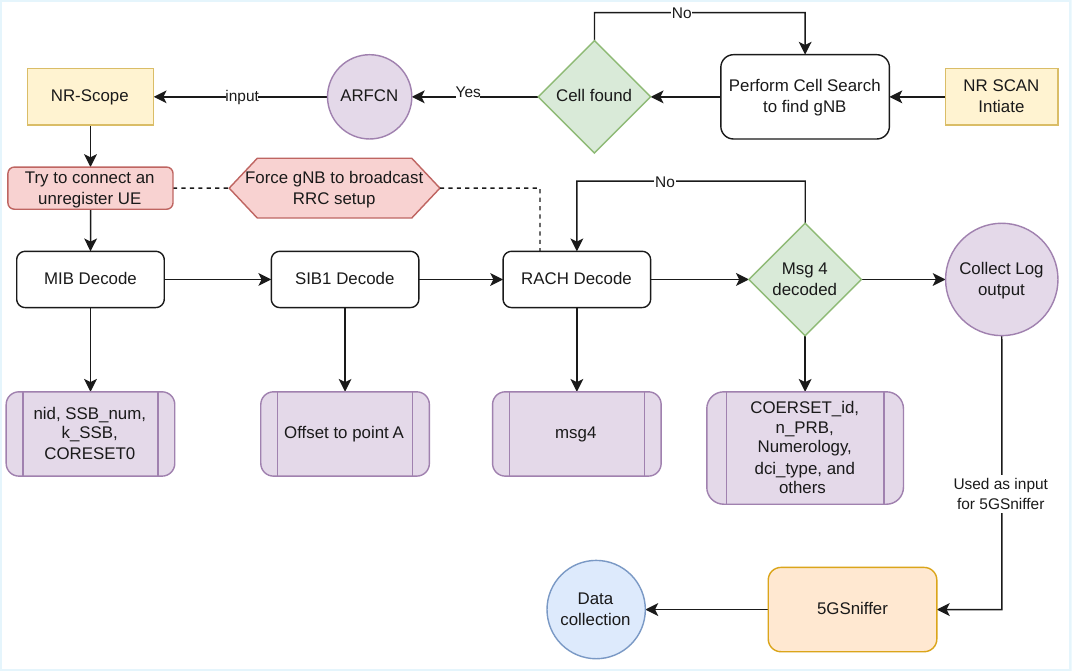}
  \caption{Data collection pipeline.} \vspace{-0.25in}
  \label{fig:data_collection}
\end{figure}

Before fingerprinting, the adversary must collect a dataset in a black-box manner and map RNTIs to physical UEs before training. Fig.~\ref{fig:data_collection} gives the full flow. Reconnaissance reuses the passive decoding chain of \S\ref{subsec:initial_attach}, performed with NR-Scope~\cite{10.1145/3680121.3697808}: the \texttt{nr-scan} module synchronizes to the SSB and recovers the ARFCN, after which MIB/SIB1 decoding yields the numerology (\texttt{subCarrierSpacingCommon}, \texttt{subcarrierSpacing}), \texttt{k\_SSB}, PCI, and \texttt{offsetToPointA}. These parameters suffice to synchronize an SDR front-end to the cell and to decode RRC with the RA-RNTI; the only steps specific to data collection are triggering the RACH to expose the UE-specific PDCCH configuration (Step~3) and configuring 5GSniffer to harvest DCI (Step~4). \vspace{0.05in}

\paragraph{\textbf{Step~3: Triggering the RACH Procedure and Observing the RRC Setup.}}
To decode UE-specific DCI, the adversary must extract the UE-specific PDCCH configuration from the \texttt{RRC~Setup} message in \texttt{msg4}. It triggers a RACH exchange by attempting to attach with an unregistered SIM; this fails authentication but still drives the RACH procedure and the transmission of \texttt{RRC~Setup}. NR-Scope decodes \texttt{msg4} and reveals the CORESET\#1 parameters: \texttt{controlResourceSetId}, \texttt{duration}, the frequency-domain resource bitmap (\texttt{frequencyDomainResources}), the CCE-to-REG mapping type (\texttt{cce-REG-MappingType}), \texttt{dci-Formats}, and \texttt{nrofCandidates}. The adversary also observes the temporary C-RNTI assigned during the exchange. Per 3GPP TS~38.321~\S5.1.4, the TC-RNTI assigned in Msg2 is promoted to C-RNTI on successful Msg4 reception without a value change. In our srsRAN-based testbed, C-RNTIs are drawn sequentially from a pool, so the adversary can infer the current assignment range after one or more RACH exchanges; commercial gNBs may randomize this assignment, in which case the inference does not apply, though nothing in the remainder depends on it. These RRC messages are observable before AS security is activated~\cite{3gpp-ts33501}, so no cryptographic keys are required.

\paragraph{\textbf{Step~4: 5GSniffer Configuration and DCI Dataset Collection.}}
Using the parameters from reconnaissance and Step~3, the adversary configures 5GSniffer~\cite{ludant20235g}, an open-source PDCCH decoder, for blind UE-specific DCI decoding. For each slot, 5GSniffer attempts CRC verification of decoded candidates across the configured C-RNTI values, recording successful decodes with their full field sets; restricting the search to the narrow range inferred in Step~3 cuts decoding complexity by several orders of magnitude. The output is a time-ordered sequence of DCI records, one per scheduling decision observed on the PDCCH.

\vspace{-0.05in}

\subsection{RNTI to UE Device Mapping}
\label{sec:cccdh_overview}

In every 5G system, the gNB assigns a temporary identifier (C-RNTI) to each UE during an RRC connection. These identifiers change across RRC sessions, so an eavesdropper cannot trivially map them to devices: whenever a UE disconnects and reconnects, whether due to mobility, inactivity timers, or RRC re-establishment, it receives a fresh C-RNTI with no visible link to the previous one. The decoded DCI records therefore appear as a stream of scheduling metadata tagged with unpredictable identifiers, and to fingerprint per device the adversary must first determine which RNTIs belong to the same physical UE.

\subsubsection{Motivation for a Graph-Theoretic Approach}
\label{sec:motivation}

The mapping problem has one hard constraint: two RNTIs active in the same time window cannot belong to the same UE, since a UE holds exactly one C-RNTI per active RRC connection~\cite{3gpp-ts38321}. This relation is pairwise, and any structure over objects (nodes) and pairwise relations (edges) is a graph. Representing each RNTI by its activity interval and joining overlapping pairs yields an \emph{interval graph}~\cite{golumbic2004algorithmic}, which is perfect; its clique number therefore equals its chromatic number, i.e., the minimum number of colors, and since one color per device is forced by the constraint, that clique number is the tightest lower bound on the device count $K$, computable in polynomial time from timestamps alone.

The conflict graph states which RNTI pairs \emph{cannot} share a device but not which non-conflicting pairs \emph{do}. Resolving that requires a second structure that is (i) derivable from timestamps alone and (ii) asymmetric in time, since reconnection is directed: an old RNTI ends and a new one begins. These properties call for a directed acyclic graph with temporal-gap edge weights, formalized next.

\subsubsection{Problem Formulation}
\label{sec:problem_formulation}

The mapping problem takes a set of observed RNTIs $\mathcal{R}=\{r_1,\ldots,r_n\}$, each with an activity interval $[\alpha(r_i),\beta(r_i)]$ from its first to last DCI timestamp, and produces a partition $\mathcal{P}=\{D_1,\ldots,D_K\}$ in which each $D_k$ is the ordered RNTI sequence of one physical UE. Neither $K$ nor any device's reconnection schedule is known in advance. We define the two graph structures used throughout:

\smallskip
\noindent\textbf{Conflict graph} $G_C=(\mathcal{R},E_C)$ encodes the one-RNTI-per-connection invariant~\cite{3gpp-ts38321}: an undirected edge $(r_i,r_j)\in E_C$ exists whenever the intervals of $r_i$ and $r_j$ overlap, i.e., $\alpha(r_i)\le\beta(r_j)$ and $\alpha(r_j)\le\beta(r_i)$.

\smallskip
\noindent\textbf{Succession DAG} $G_S=(\mathcal{R},E_S,w)$ encodes candidate same-device transitions: a directed edge $(r_i\!\to\!r_j)\in E_S$ exists iff $\beta(r_i)<\alpha(r_j)$ and $(r_i,r_j)\notin E_C$. Because time is irreversible, $G_S$ is acyclic. Each edge carries a squared-gap cost
\[
  w(r_i,r_j)\;=\;\bigl(\alpha(r_j)-\beta(r_i)\bigr)^2 ,
\]
encoding that a reconnecting device minimizes its own reconnection delay; squaring penalizes long silences superlinearly (a $10\,$s gap costs $100\times$ as much as a $1\,$s gap), matching the physical bound that the delay lies between the RRC procedure time and the inactivity timer.

\smallskip
\noindent\textbf{Conflict-Constrained Path Cover (CCPC).} Partition $\mathcal{R}$ into vertex-disjoint directed paths (chains) in $G_S$ that (i)~cover every $r\in\mathcal{R}$ exactly once and (ii)~place no two $G_C$-adjacent RNTIs on one path (enforced by construction, since $E_S$ excludes conflicting pairs), while minimizing
\begin{equation}
J(\mathcal{P}) \;=\; \sum_{(r_i\to r_j)\in\mathcal{P}} w(r_i,r_j) \;+\; c\,K(\mathcal{P}),
\label{eq:cccp_objective}
\end{equation}
where $K(\mathcal{P})$ is the number of chains and $c>0$ is the cost of opening a new chain, equivalently of leaving a node without a matched successor. The penalty $c$ is essential. Without it, every $w\ge 0$ makes the empty cover ($K=n$, $J=0$) trivially optimal, so a plain minimum-weight matching would return the degenerate solution that treats each RNTI as its own device. With it, linking $r_i\to r_j$ is preferred over opening a new chain exactly when $w(r_i,r_j)<c$, i.e., when the reconnection gap is below $\tau_{\max}=\sqrt{c}$. We treat $\tau_{\max}$ as a confidence threshold on the reconnection gap rather than a hard physical bound: sub-$\tau_{\max}$ gaps (e.g., RRC re-establishment or brief idle) are strong evidence of same-device continuity, whereas longer gaps are uninformative on their own and are deferred to Phase~3.

\smallskip
\noindent\textbf{Reduction to assignment.} Minimizing \eqref{eq:cccp_objective} reduces to a bipartite assignment. Split each $r$ into a successor port $r^L$ and a predecessor port $r^R$, and for every $(r_i\to r_j)\in E_S$ with $w(r_i,r_j)<c$ add an edge $(r_i^L,r_j^R)$ with reward $c-w(r_i,r_j)>0$. Each matched edge merges two chains and lowers $J$ by exactly $c-w$, since $K$ drops by one ($c$ saved) at an added link cost $w$. Minimizing $J$ is therefore equivalent to a \emph{maximum-weight} bipartite matching on these rewards, solved by the Hungarian algorithm~\cite{kuhn1955hungarian} in $O(n^3)$ time. Because only sub-$\tau_{\max}$ links are admitted, genuine but long reconnections are deliberately left unlinked, which is what makes the resulting chain count an \emph{upper} bound on $K$ that Phase~3 then tightens.

\subsubsection{Proposed Solution}
\label{sec:rnti_solution}

We solve CCPC with \emph{CCCD}, which uses only \emph{timestamp}, \emph{RNTI}, and \emph{direction} (UL/DL); these depend on \emph{whether} and \emph{when} a device was scheduled, not \emph{how}, making them invariant to the gNB's resource-allocation decisions. CCCD has three phases:

\begin{enumerate}[leftmargin=1.8em, topsep=2pt, itemsep=1pt]
  \item \textbf{Conflict Detection} (\S\ref{sec:conflict_graph}): identify simultaneously active RNTIs that cannot share a device.
  \item \textbf{Succession Scoring} (\S\ref{sec:succession_dag}): among non-conflicting RNTIs, score how likely one is the direct temporal successor of another.
  \item \textbf{Chain Cover and Model Selection} (\S\ref{sec:chain_cover}): stitch RNTIs into per-device chains via the matching above, then estimate $K$ with the Bayesian Information Criterion (BIC).
\end{enumerate}

\begin{figure}[t] \vspace{-0.05in}
  \centering
  \includegraphics[width=0.9\columnwidth]{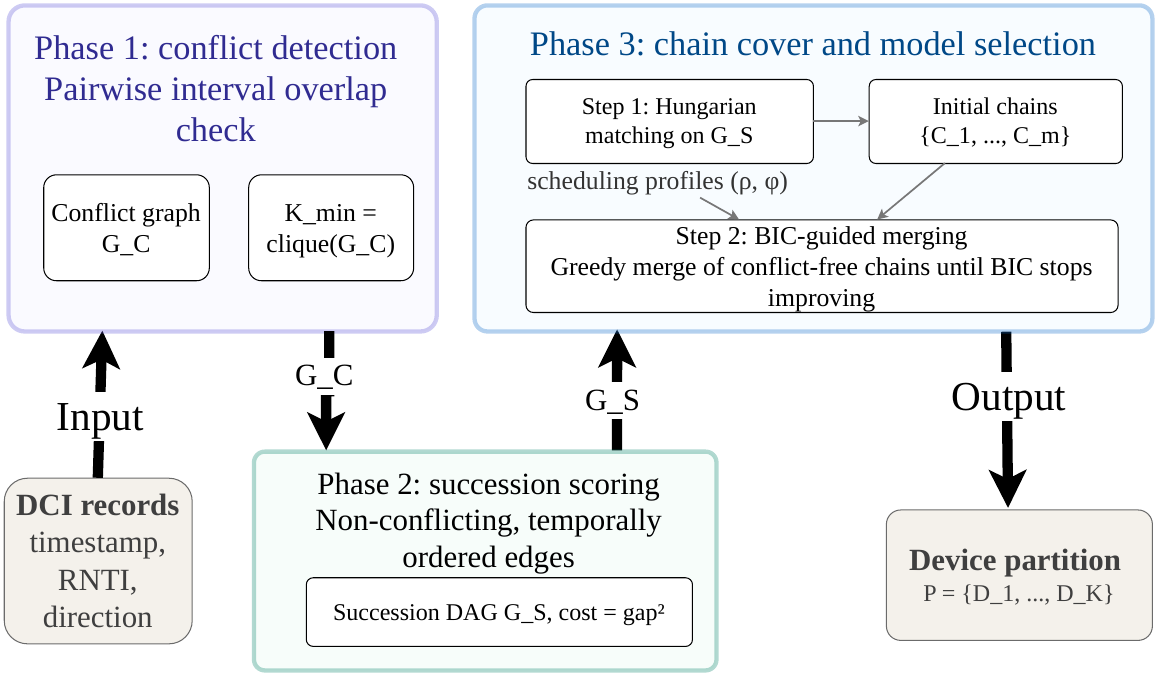}
  \caption{Dataflow of CCCD.} \vspace{-0.25in}
  \label{fig:cccd_phases}
\end{figure}

\noindent
Figure~\ref{fig:cccd_phases} summarizes the dataflow, and Algorithm~\ref{alg:cccd_revised} gives the consolidated procedure.

\subsubsection{Phase~1: Conflict Detection}
\label{sec:conflict_graph}

Phase~1 instantiates $G_C$ (\S\ref{sec:problem_formulation}) via the $O(n^2)$ pairwise interval-overlap test and reports the lower bound $K_{\min}=\omega(G_C)$ at no additional cost. Since $n$, the number of distinct RNTIs in a session, is on the order of tens to hundreds, this step is negligible in time.

\subsubsection{Phase~2: Temporal Succession Scoring}
\label{sec:succession_dag}

Phase~2 instantiates $G_S$ (\S\ref{sec:problem_formulation}), admitting an edge $(r_i\!\to\!r_j)$ only when the pair is conflict-free and forward in time ($\beta(r_i)<\alpha(r_j)$) and weighting it by the squared gap $w(r_i,r_j)$. The first condition ensures conflict feasibility, the second enforces causality, and the weight encodes the preference for short reconnection delays defined above.

\subsubsection{Phase~3: Chain Cover and Model Selection}
\label{sec:chain_cover}

Phase~3 converts $G_S$ into per-device RNTI sequences and determines the device count.

\smallskip
\noindent\textbf{Step~1: Minimum-cost chain cover.}
A \emph{chain} is a directed path in $G_S$ (e.g., $r_3\!\to\!r_7\!\to\!r_{12}$ means one device used three successive RNTIs), and a \emph{chain cover} assigns every RNTI to exactly one chain. We compute the cover that minimizes $J$ \eqref{eq:cccp_objective} by the maximum-weight matching of \S\ref{sec:problem_formulation}, then convert matched successor links into vertex-disjoint chains. Because a link is taken only when its squared gap is below $c$, chains break wherever a reconnection gap exceeds $\tau_{\max}=\sqrt{c}$, so a device that went idle for longer than $\tau_{\max}$ is split across several chains. This deliberate splitting is what makes the resulting count $m$ an \emph{upper} bound on $K$; Step~2 then merges the over-split chains using their temporal profiles rather than the gap.

\smallskip
\noindent\textbf{Step~2: BIC-guided merging.}
The initial count $m$ upper-bounds $K$. To decide whether two chains separated by a gap too long for Phase~2 to link cheaply belong to the same device, we use the Bayesian Information Criterion (BIC)~\cite{schwarz1978estimating}, which balances fit against complexity. For each RNTI~$r$ with $\beta(r)>\alpha(r)$ we form a two-dimensional temporal profile
\begin{equation}
\mathbf{t}(r)=\bigl(\rho(r),\,\phi(r)\bigr),
\label{eq:temporal_profile}
\end{equation}
where $\rho(r)=n(r)/(\beta(r)-\alpha(r))$ is the scheduling density and $\phi(r)=n_{\mathrm{UL}}(r)/n(r)$ is the uplink fraction; single-observation RNTIs are retained as singletons. Each device cluster is a diagonal bivariate Gaussian over these profiles, and BIC accepts a merge only when the combined model scores strictly better than the two separate models. The procedure is greedy: at each iteration the conflict-free, merge-eligible pair with the largest BIC improvement is merged (a merge is conflict-free if no RNTI in one chain conflicts with any in the other), and iteration stops when no merge improves BIC. The remaining chain count is the estimated $K$, obtained without specifying $K$ in advance.


\begin{algorithm}[t] 
\caption{Conflict-Constrained Chain Decomposition}
\label{alg:cccd_revised}
\footnotesize
\begin{algorithmic}[1]
\Require DCI records $\mathcal{X}=\{(t, r, d)\}$ ($t$ timestamp, $r$ RNTI, $d\in\{UL,DL\}$); new-chain penalty $c=\tau_{\max}^2$
\Ensure Device partition $\mathcal{P}=\{D_1,\ldots,D_K\}$ and estimated device count $K$

\State Aggregate $\mathcal{X}$ by RNTI to obtain $\mathcal{R}$
\ForAll{$r\in\mathcal{R}$}
    \State $\alpha(r)\gets \min\{t:(t,r,d)\in\mathcal{X}\}$; \quad $\beta(r)\gets \max\{t:(t,r,d)\in\mathcal{X}\}$
    \State $n(r)\gets |\{(t,r,d)\in\mathcal{X}\}|$; \quad $n_{UL}(r)\gets |\{(t,r,UL)\in\mathcal{X}\}|$
\EndFor

\Statex \textit{// Phase 1: hard conflict constraints}
\State $G_C\gets(\mathcal{R},\emptyset)$
\ForAll{unordered pairs $\{r_i,r_j\}\subset\mathcal{R}$}
    \If{$\alpha(r_i)\le\beta(r_j)$ \textbf{and} $\alpha(r_j)\le\beta(r_i)$}
        \State Add conflict edge $\{r_i,r_j\}$ to $G_C$
    \EndIf
\EndFor
\State $K_{\min}\gets \omega(G_C)$ \Comment{minimum feasible number of physical UEs}

\Statex \textit{// Phase 2: feasible temporal transitions}
\State $G_S\gets(\mathcal{R},\emptyset)$
\ForAll{ordered pairs $(r_i,r_j)$ with $r_i\ne r_j$}
    \If{$\{r_i,r_j\}\notin E_C$ \textbf{and} $\beta(r_i)<\alpha(r_j)$}
        \State $w_{ij}\gets (\alpha(r_j)-\beta(r_i))^2$
        \State Add edge $(r_i\rightarrow r_j)$ with cost $w_{ij}$ to $G_S$
    \EndIf
\EndFor

\Statex \textit{// Phase 3a: initial chain cover (new-chain penalty $c$)}
\State Split each $r\in\mathcal{R}$ into left and right copies $(r^L,r^R)$
\State Build a bipartite graph: for every $(r_i\rightarrow r_j)\in E_S$ with $w_{ij}<c$, add edge $(r_i^L,r_j^R)$ with reward $c-w_{ij}>0$
\State $M\gets$ maximum-weight matching on this bipartite graph \Comment{minimizes $J$ in \eqref{eq:cccp_objective}}
\State Convert matched successor links in $M$ into vertex-disjoint chains $\mathcal{C}=\{C_1,\,\ldots,\,C_m\}$, $m\ge K$

\Statex \textit{// Phase 3b: BIC-guided model selection}
\ForAll{$r\in\mathcal{R}$ with $\beta(r)>\alpha(r)$}
    \State $\rho(r)\gets n(r)/(\beta(r)-\alpha(r))$; \quad $\phi(r)\gets n_{UL}(r)/n(r)$
    \State $\mathbf{t}(r)\gets(\rho(r),\phi(r))$
\EndFor
\Repeat
    \State Among all conflict-free chain pairs, find the pair $(C_a,C_b)$ with the largest BIC improvement after merging
    \If{the best merge strictly improves BIC}
        \State $\mathcal{C}\gets(\mathcal{C}\setminus\{C_a,C_b\})\cup\{C_a\cup C_b\}$
    \EndIf
\Until{no conflict-free merge improves BIC}
\State $\mathcal{P}\gets\mathcal{C}$; \quad $K\gets|\mathcal{P}|$
\State \Return $\mathcal{P},K$
\end{algorithmic}
\end{algorithm}\vspace{0.15in}
\vspace{-0.05in}

\subsection{Trace Reconstruction and Preprocessing}
\label{subsec:reconstruction}

After RNTI-to-device mapping, the raw DCI stream is separated into client-specific scheduling traces. Each trace is represented as an ordered sequence of decoded scheduling records, each containing a timestamp, transmission direction, transport block size, and physical resource block allocation. We then preprocess each per-device trace before extracting temporal fingerprints.

First, we remove low-activity or irrelevant scheduling records using a transport block size filter. The remaining records are grouped into uplink and downlink bursts based on temporal proximity. These bursts provide the basis for identifying the FL round-cycle structure, in which each round typically includes downlink activity associated with global model delivery and uplink activity associated with client update transmission.

However, passive over-the-air DCI collection is inherently incomplete. A sniffer may miss scheduling records because of decoding failures, temporary synchronization loss, or weak signal conditions. Missing records can introduce artificial gaps in the client-specific trace and distort burst boundaries, which directly affects View R and View L because they rely on cadence and round-to-round temporal structure. To reduce this effect, \textsc{Flint} applies a missing-grant reconstruction step to each per-device trace. The goal is not to recover packet contents or user-plane data, but to restore the temporal continuity of the scheduling trace by identifying abnormal gaps and imputing likely missing scheduling activity from neighboring observations. Algorithm~\ref{alg:reconstruction} summarizes this reconstruction process.

\begin{algorithm}[t]
\caption{Data Reconstruction (Packet-Loss Imputation)}
\label{alg:reconstruction}
\begin{algorithmic}[1]
\Require Raw grants $G$; $\tau_{ul}{=}300$, $\tau_{dl}{=}150$; $\delta{=}1.0$s; $m$; $\lambda{=}1.8$; $\kappa$
\Ensure Imputed grant set $\tilde{G}$

\State $G \gets \{g \in G : (g.\text{dir}{=}\text{UL} \wedge g.\text{tb}{\ge}\tau_{ul}) \vee
       (g.\text{dir}{=}\text{DL} \wedge g.\text{tb}{\ge}\tau_{dl})\}$
\State $B_{ul}, B_{dl} \gets \textsc{DetectBursts}(G,\delta)$; discard bursts with ${<}\,m$ grants
\State $B_{dl} \gets \textsc{KMeansHeightFilter}(B_{dl}, k{=}2)$
\State $\mu_s \gets \mathrm{med}(|B_{ul}|)$; $\;\rho \gets \mathrm{med}(\text{DL-to-DL spacing})$
\State $\mathcal{T}_{ul}, \mathcal{T}_{dl} \gets$ empirical TB-size pools from $B_{ul}, B_{dl}$
\State $S \gets \emptyset$
\For{each consecutive pair $(d_i, d_{i+1})$ in $B_{dl}$}
    \If{no UL burst precedes $d_i$}
        \State $S \gets S \cup \textsc{SampleBurst}(\text{UL}, \mu_s, \mathcal{T}_{ul}, d_i)$
    \EndIf
    \If{$d_{i+1}.\text{start} - d_i.\text{end} > \lambda\rho$}
        \State $n \gets \min(\lfloor(d_{i+1}.\text{start} - d_i.\text{end})/\rho\rceil - 1,\; \kappa)$
        \State $S \gets S \cup \bigcup_{r=1}^{n} \textsc{SampleRound}(\mathcal{T}_{ul}, \mathcal{T}_{dl}, \mu_s, \rho)$
    \EndIf
\EndFor
\State $\tilde{G} \gets (G \cup S)$ with UL grants inside DL windows removed
\State \Return $\tilde{G}$
\end{algorithmic}
\end{algorithm}

\subsection{Fingerprinting Framework}
\label{subsec:framework}

The fingerprinting framework takes a cleaned per-device DCI stream and produces an architecture label. In Figure~\ref {fig:system}, we see that it consists of four stages: segmentation into observation windows, extraction of three complementary temporal views, late fusion of view-level predictions, and the final fingerprint decision.

\subsubsection{Segmentation}
\label{subsubsec:segmentation}

A passive observer may begin monitoring a client at any point during the training session. Therefore, \textsc{Flint} operates on fixed-duration observation windows rather than complete training sessions. We slide a window of length $w$ over each client's DCI stream and treat each window as an independent attack observation. We set $w=300,s$. This duration provides enough time to capture multiple federated training rounds while remaining practical for passive monitoring.

Within each window, \textsc{Flint} identifies round-cycle structure from the alternation between uplink and downlink activity. This produces a sequence of round-level segments used by the rhythm and sequential views. Windows shorter than $w$ at the end of a stream are retained as independent observations only if they contain sufficient activity.

\subsubsection{Three Temporal Views}
\label{subsubsec:views}

Simple statistics derived from transport block sizes and resource block allocations do not reliably distinguish model families when their overall communication patterns are similar. \textsc{Flint} therefore represents each observation window through three complementary temporal views. Each view is trained independently and produces a class-probability vector. The separate views allow the framework to capture different aspects of training behavior and provide robustness when one representation is less informative for a particular window.

\paragraph{View W: Multi-Resolution Energy.}
View W captures how scheduling activity is distributed across multiple time scales. As summarized in Algorithm~\ref{alg:view_w}, we first convert each observation window into a regularly sampled activity signal based on transport block sizes or resource allocation activity. We then apply a discrete wavelet transform to decompose the signal into components at successively coarser resolutions. The energy at each resolution level is summarized and normalized to obtain a scale-aware descriptor of temporal activity. This view captures whether activity appears as sharp short-lived bursts, slower fluctuations, or a mixture of both. The resulting feature vector is classified using XGBoost, producing the probability vector $\mathbf{p}_W$.

\begin{algorithm}[t]
\caption{View W: Wavelet Multi-Resolution Energy}
\label{alg:view_w}
\begin{algorithmic}[1]
\Require Window grants $G_w$; number of slots $N{=}128$; wavelet levels $L{=}6$;
         XGBoost classifier $h_W$
\Ensure View probability vector $\mathbf{p}_W \in \Delta^2$

\State $x \gets \textsc{ActivitySignal}(G_w, N)$
    \Comment{$x_j = \sum_{g \in \text{slot } j} g.\text{tb}$}
\State $\mathbf{e} \gets [\,]$

\For{$\ell = 0$ \textbf{to} $L - 1$}
    \State $a_i \gets (x_{2i} + x_{2i+1})/\sqrt{2}$
        \Comment{Haar approximation}
    \State $d_i \gets (x_{2i} - x_{2i+1})/\sqrt{2}$
        \Comment{Haar detail}
    \State Append $e_\ell \gets \sum_i d_i^2$ to $\mathbf{e}$
        \Comment{Detail energy at scale $\ell$}
    \State $x \gets a$
        \Comment{Recurse on approximation}
\EndFor

\State Append $e_L \gets \sum_i x_i^2$ to $\mathbf{e}$
    \Comment{Residual approximation energy}
\State $\mathbf{x}_W \gets \mathbf{e} \,/\, \sum_{\ell=0}^{L} e_\ell$
    \Comment{Normalize to energy distribution}
\State $\mathbf{p}_W \gets h_W(\mathbf{x}_W)$
\State \Return $\mathbf{p}_W$
\end{algorithmic}
\end{algorithm}

\paragraph{View R: Cadence and Regularity.}
View R captures the periodicity and regularity of the federated training cycle. As shown in Algorithm~\ref{alg:view_r}, we use the round-cycle boundaries identified during segmentation to extract statistical descriptors of the round structure, including timing, uplink activity, downlink activity, and phase durations. We also compute spectral descriptors from round-level sequences to measure how strongly the traffic follows a regular cadence. A stable training loop tends to produce more regular round timing, while irregular computation or communication behavior produces a more diffuse temporal pattern. These features are classified using XGBoost, producing the probability vector $\mathbf{p}_R$.

\begin{algorithm}[t]
\caption{View R: Round-Cadence Regularity and Spectral Features}
\label{alg:view_r}
\begin{algorithmic}[1]
\Require Window grants $G_w$; XGBoost classifier $h_R$
\Ensure $\mathbf{p}_R \in \Delta^2$
\State $\mathcal{R} \gets \textsc{ExtractRounds}(G_w)$
\State Form sequences: gaps $\mathbf{g}$, UL sizes $\mathbf{u}$, DL sizes $\mathbf{d}$, durations $\boldsymbol{\delta}$
\For{each $s \in \{\mathbf{g}, \mathbf{u}, \mathbf{d}, \boldsymbol{\delta}\}$}
    \State Append $\mu(s),\, \sigma(s),\, \mathrm{slope}(s),\, \mathrm{AC}_1(s)$ to $\mathbf{x}_{\text{stat}}$
\EndFor
\For{each $s \in \{\mathbf{g}, \mathbf{u}\}$}
    \State $p(f) \gets |\textsc{FFT}(s - \mu(s))|^2 \,/\, (\textstyle\sum_f |\textsc{FFT}(s-\mu(s))|^2 + \epsilon)$
    \State Append $\arg\max_f p(f),\; {-}\sum_f p(f)\log p(f),\; \max_f p(f)$ to $\mathbf{x}_{\text{spec}}$
\EndFor
\State $\mathbf{x}_R \gets [\,|\mathcal{R}| \;\|\; \mathbf{x}_{\text{stat}} \;\|\; \mathbf{x}_{\text{spec}}\,]$
\State $\mathbf{p}_R \gets h_R(\mathbf{x}_R)$
\State \Return $\mathbf{p}_R$
\end{algorithmic}
\end{algorithm}

\paragraph{View L: Sequential Evolution.}
View L models the round-to-round evolution of scheduling activity. As described in Algorithm~\ref{alg:view_l}, each window is represented as an ordered sequence of per-round descriptors and used to train a bidirectional Long Short-Term Memory network. The Bi-LSTM captures dependencies between consecutive rounds that are not preserved by aggregate statistical features. This view is sensitive to how the training and update pattern changes over the observation window. At inference time, the trained Bi-LSTM produces the probability vector $\mathbf{p}_L$ for each observation window.

\begin{algorithm}[t]
\caption{View L: Round-Sequence BiLSTM Training}
\label{alg:view_l}
\begin{algorithmic}[1]
\Require $\{(G_w, y_w)\}_{w=1}^{M}$; class weights $w_c \propto 1/n_c$; epochs $E$
\Ensure Trained BiLSTM $f_\theta$
\State $\mathbf{R}_w \gets \textsc{ExtractRounds}(G_w)$, z-scored by $(\boldsymbol{\mu}_f, \boldsymbol{\sigma}_f)$ $\forall\, w$
\State Initialize 2-layer BiLSTM $f_\theta$ with softmax over $\mathcal{C} = \{\text{CNN},\text{RNN},\text{Trans}\}$
\State Optimize $\theta$ via Adam ($E$ epochs):
\[
\mathcal{L}(\theta) = -\frac{1}{M}\sum_{w=1}^{M}\sum_{c \,\in\, \mathcal{C}}
w_c\; y_{w,c}\;\log f_\theta(\mathbf{R}_w)_c
\]
\State \Return $f_\theta$
\Statex \textit{Inference:} $\mathbf{p}_L \gets f_\theta(\mathbf{R}_w) \in \Delta^{|\mathcal{C}|-1}$
\end{algorithmic}
\end{algorithm}

\subsubsection{Late Fusion}
\label{subsubsec:fusion}

The three views provide complementary evidence about the client's architecture. Instead of concatenating raw feature vectors, \textsc{Flint} performs late fusion over the view-level probability vectors $\mathbf{p}_W$, $\mathbf{p}_R$, and $\mathbf{p}_L$. Specifically, it forms a stacked probability representation \vspace{-0.15in}
\[
\mathbf{z} = [\mathbf{p}_W \,\|\, \mathbf{p}_R \,\|\, \mathbf{p}_L],
\]
where $|$ denotes concatenation. A meta-classifier $g*\phi$, implemented as logistic regression, maps this stacked representation to a fused class-probability vector: \vspace{-0.1in}
\[
\mathbf{p}*F = g*\phi(\mathbf{z}).
\]
The meta-classifier is trained using view-level predictions generated from the training folds. This probability-level fusion keeps the views modular, allows each view to be trained and inspected independently, and improves robustness when one view is degraded by missing observations or weak temporal structure and we called this fusion "stacking" fusion.

\subsubsection{Fingerprint Decision}
\label{subsubsec:decision}

In the closed-world setting, the predicted architecture is the class with the highest fused probability:
\[
\hat{y} = \arg\max_{c \in {\text{CNN},\text{RNN},\text{Transformer}}} \mathbf{p}_F(c).
\]
In the open-world setting, \textsc{Flint} rejects low-confidence windows as unknown:
\[
\hat{y} =
\begin{cases}
\emph{Others}, & \text{if } \max_c \mathbf{p}_F(c) < \eta,\\
\arg\max_c \mathbf{p}_F(c), & \text{otherwise}.
\end{cases}
\]
The final client-level fingerprint is obtained by aggregating predictions across multiple windows from the same client.
\vspace{-0.05in}
\vspace{-0.05in}

\section{Evaluation Result and Analysis}
\label{sec:evaluation}


\subsection{Experimental Setup and Metrics}
\label{sec:setup}

\noindent\textbf{Testbed.}
We evaluate \textsc{Flint} on a real 5G federated learning testbed. The testbed consists of heterogeneous FL client devices connected through a 5G cell served by a gNodeB. Each client trains deep learning model locally and exchanges model updates with a federated server through the 5G network. A passive SDR-based sniffer collects PDCCH scheduling information and decodes DCI records without interacting with the network, the gNodeB, or the FL clients.

\noindent\textbf{Architecture families.}
We evaluate three architecture families: convolutional neural networks (CNNs), recurrent neural networks (RNNs), and Transformers. Each family contains multiple concrete model variants so that the classification target is the architecture family rather than a specific model instance. For open-world experiments, we additionally consider unseen model variants and non-FL background traffic as \emph{Others}.

\noindent\textbf{Evaluation settings.}
We evaluate \textsc{Flint} under closed-world and open-world settings. In the closed-world setting, each observation window belongs to one of the three known FL architecture families, and the task is three-way classification among CNN, RNN, and Transformer. In the open-world setting, the model is trained on known FL families and evaluated on unseen model variants or background traffic, requiring the classifier to reject unknown windows as \emph{Others}.

\noindent\textbf{Observation windows.}
Unless otherwise stated, we use fixed observation windows of $w=300$,s. This window length captures multiple FL round cycles while remaining practical for passive monitoring. Section~\ref{sec:ablation_window} studies the effect of different observation window sizes.

\noindent\textbf{Training protocol.}
For closed-world evaluation, we use stratified 5-fold cross-validation with balanced class pools. Balancing prevents the classifier from favoring the majority architecture family and ensures that performance reflects per-class separability rather than dataset composition. For each fold, feature extraction, view-level classifiers, and the fusion meta-classifier are trained only on the training split and evaluated on the held-out split.

\noindent\textbf{Metrics.}
We report per-class precision, recall, and F1-score, as well as macro-averaged precision, recall, and F1-score. Unless otherwise stated, we report mean and standard deviation across cross-validation folds. For open-world rejection, we also report unknown-detection performance and false-positive behavior for the \emph{Others} class.
\vspace{-0.05in}

\subsection{RNTI Mapping Evaluation}
\label{sec:rnti_mapping_eval}

\begin{table}[b]
\centering
\renewcommand{\arraystretch}{1.45}
\setlength{\tabcolsep}{6pt}
\caption{Confusion matrix of CCCD RNTI-to-UE mapping (row-normalized).
Rows are ground-truth UEs. CCCD assigns every RNTI to the correct device.}
\label{tab:cccd_confusion}
\begin{tabular}{c c ccc}
\toprule
 & & \multicolumn{3}{c}{\textbf{Predicted (aligned)}} \\
\cmidrule(lr){3-5}
 & & UE$_A$ & UE$_B$ & UE$_C$ \\
\midrule
\multirow{3}{*}{\rotatebox[origin=c]{90}{\small\textbf{GT}}}
 & UE$_A$ & \textbf{1.00} & 0.00 & 0.00 \\
 & UE$_B$ & 0.00 & \textbf{1.00} & 0.00 \\
 & UE$_C$ & 0.00 & 0.00 & \textbf{1.00} \\
\bottomrule
\end{tabular}
\end{table}

Here, we evaluate CCCD on a $613\,$s capture of $39{,}669$ DCI records carried by
eight C-RNTIs from three UEs, using only timestamp, RNTI, and direction with a
reconnection window $\tau_{\max}=500\,$ms. After aligning each
recovered chain to a ground-truth device for scoring only, CCCD estimates
$\hat{K}=3$ directly from the data and recovers every chain exactly. Therefore, the confusion matrix is diagonal (Table~\ref{tab:cccd_confusion}) at
$100\%$ RNTI accuracy ($8/8$) and $100\%$ record accuracy ($39{,}669/39{,}669$).\vspace{-0.05in}


\subsection{Fingerprinting Performance}
\label{sec:fingerprint_performance}

\noindent\textbf{Closed-world performance.}
Table~\ref{tab:main_closed_world} reports the closed-world performance of \textsc{Flint} using the full three-view configuration with stacking-based late fusion at $w=300$,s. \textsc{Flint} achieves a macro F1-score of $0.930 \pm 0.021$ across CNN, RNN, and Transformer families. RNNs are identified with the highest F1-score, reflecting their distinctive temporal scheduling behavior. CNNs and Transformers are difficult to distinguish because they can exhibit similar aggregate communication patterns, yet \textsc{Flint} still reliably identifies both families.

\begin{table}[t]
\centering
\caption{Main closed-world performance at $w=300$ s, stacking fusion.}
\label{tab:main_closed_world}
\begin{tabular}{lccc}
\toprule
Class & Precision & Recall & F1 \\
\midrule
CNN & 0.921 $\pm$ 0.047 & 0.871 $\pm$ 0.030 & 0.895 $\pm$ 0.031 \\
RNN & 0.963 $\pm$ 0.030 & 0.994 $\pm$ 0.013 & 0.978 $\pm$ 0.019 \\
Transformer & 0.909 $\pm$ 0.018 & 0.929 $\pm$ 0.040 & 0.918 $\pm$ 0.016 \\
\midrule
\textbf{Macro} & \textbf{0.931} & \textbf{0.931} & \textbf{0.930 $\pm$ 0.021} \\
\bottomrule
\end{tabular}
\end{table}

\noindent\textbf{Insight.}
\textit{These results confirm that the 5G control channel leaks more information than it appears to at first glance. Although the adversary does not see packets or payloads, the timing and size of scheduling grants still reflect how different model families behave during federated training. The high F1-score shows that FLINT can recover this hidden structure from coarse PHY-layer observations.}

\begin{figure}[t]
\centering
\includegraphics[width=0.7\columnwidth]{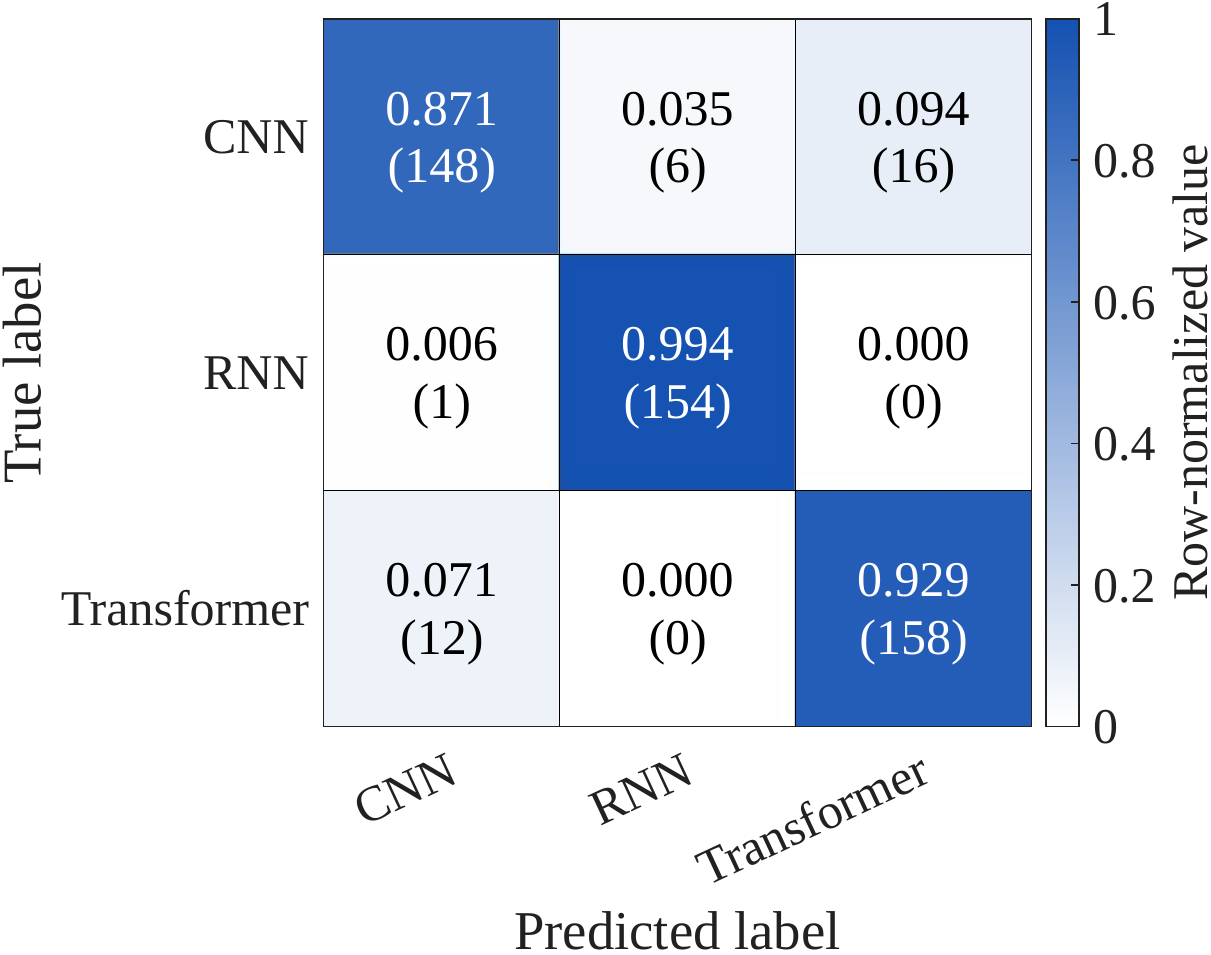}
\caption{Normalized confusion matrix for closed-world fingerprinting at $w=300$,s using stacking fusion. Rows indicate true architecture families and columns indicate predicted families.}
\vspace{-0.05 in}
\label{fig:confusion_closed}
\end{figure}

\noindent\textbf{Confusion analysis.}
Figure~\ref{fig:confusion_closed} shows the normalized confusion matrix for the closed-world setting at $w=300$,s using stacking fusion. RNN samples are almost perfectly separated from the other families, with only one RNN sample misclassified as CNN. Most errors occur between CNN and Transformer, as expected, because these families tend to exhibit more similar aggregate scheduling behavior than RNNs. Nevertheless, \textsc{Flint} correctly identifies 87.1\% of CNN samples and 92.9\% of Transformer samples, showing that temporal evidence from the three views substantially reduces CNN-Transformer confusion.

\noindent\textbf{Insight.}
\textit{The confusion matrix shows that \textsc{FLINT} separates RNNs reliably, while most remaining errors occur between CNNs and Transformers. This indicates that the main challenge is distinguishing architectures with similar communication volume, where temporal structure becomes the key differentiator.}

\noindent\textbf{Attack reliability.}
The per-class precision values in Table~\ref{tab:main_closed_world} show that \textsc{Flint} does not simply produce high recall with many false positives. For example, the Transformer class achieves a precision of 0.908 and a recall of 0.929. A passive adversary can therefore act on high-confidence identifications rather than relying on noisy guesses.

\noindent\textbf{Insight.}
\textit{For the fingerprinting attack, precision is as important as recall. High precision indicate that when \textsc{Flint} identifies a model family, the adversary can use that information to support downstream decisions such as architecture-specific probing, resource targeting, or selective monitoring.}
\vspace{-0.05in}


\subsection{Ablation Studies}
\label{sec:ablation}

We conduct ablation studies to explain why \textsc{Flint} works and which components are necessary. The studies answer five questions: whether simple communication statistics are sufficient, whether the three temporal views provide complementary information, what is the right fusion strategy, whether data-loss imputation matters, and how the observation window affects performance.

\noindent\textbf{Baseline and imputation analysis.}
Table~\ref{tab:baseline_imputation} compares the FLARE baseline with its statistical features to our \textsc{Flint} and also isolates the effect of data-loss imputation. Also, FLARE relies on packet-level observations, such as packet sizes, directions, timing, and inter-arrival statistics, which are available from encrypted Wi-Fi traffic but are not visible to a passive 5G PHY-layer observer. In contrast, \textsc{Flint} uses only decoded PDCCH/DCI scheduling records, including transport block sizes, transmission directions, timestamps, and physical resource block allocations. Therefore, we adapt the FLARE baseline by replacing packet-level quantities with their closest observable PHY-layer counterparts, such as TBS, direction, and timing-derived statistics. The \textsc{Flint} w/o Imp. row removes data-loss imputation from the proposed temporal multi-view pipeline while keeping the rest of the \textsc{Flint} framework unchanged. The FLARE baseline achieves a macro F1 of $0.834$, indicating that statistical summaries of PHY-layer scheduling records provide some architectural signal but are not sufficient. Adding imputation improves the FLARE  baseline to $0.877$, indicating that missing-grant reconstruction helps recover useful temporal continuity. However, the full \textsc{Flint} pipeline achieves the best macro F1 of $0.930$, showing that imputation and temporal multi-view modeling are both necessary for reliable architecture fingerprinting.

\begin{table}[t]
\centering
\caption{Baseline comparison and imputation ablation at $w=300$ s. Entries are F1-scores over 5-fold CV.}
\label{tab:baseline_imputation}
\scriptsize
\setlength{\tabcolsep}{1.7pt}
\begin{tabular}{@{}lcccc@{}}
\toprule
Method & CNN & RNN & Trans. & Macro \\
\midrule
FLARE-style & $0.825{\pm}0.058$ & $0.880{\pm}0.099$ & $0.798{\pm}0.058$ & $0.834{\pm}0.047$ \\
FLARE-style+Imp. & $0.819{\pm}0.035$ & $0.947{\pm}0.032$ & $0.865{\pm}0.077$ & $0.877{\pm}0.033$ \\
\textsc{Flint} w/o Imp. & $0.814{\pm}0.066$ & $0.884{\pm}0.059$ & $0.887{\pm}0.067$ & $0.862{\pm}0.054$ \\
\textbf{\textsc{Flint}} & $\mathbf{0.895{\pm}0.031}$ & $\mathbf{0.978{\pm}0.019}$ & $\mathbf{0.918{\pm}0.016}$ & $\mathbf{0.930{\pm}0.021}$ \\
\bottomrule
\end{tabular}
\end{table}

\noindent\textbf{Insight.}
\textit{The baseline comparison shows that simple PHY-level statistics contain useful signal, but they are not sufficient. FLINT performs better because it combines missing-grant reconstruction with temporal multi-view modeling.}

\subsubsection{Multi-View Contribution}
\label{sec:ablation_views}

Table~\ref{tab:view_ablation_detailed} evaluates all combinations of View W, View R, and View L. Single views capture useful but incomplete evidence. Pairwise combinations improve performance, and the full W+R+L configuration achieves the highest macro F1.

\begin{table}[t]
\centering
\caption{View ablation and fusion strategy comparison at $w=300$ s.}
\label{tab:view_ablation_detailed}
\resizebox{\columnwidth}{!}{%
\begin{tabular}{lcccc}
\toprule
View & CNN F1 & RNN F1 & Trans. F1 & Macro F1 \\
\midrule
W & 0.833 $\pm$ 0.020 & 0.947 $\pm$ 0.026 & 0.883 $\pm$ 0.022 & 0.888 $\pm$ 0.004 \\
R & 0.864 $\pm$ 0.043 & 0.981 $\pm$ 0.015 & 0.881 $\pm$ 0.041 & 0.909 $\pm$ 0.030 \\
L & 0.766 $\pm$ 0.049 & 0.911 $\pm$ 0.037 & 0.853 $\pm$ 0.037 & 0.843 $\pm$ 0.036 \\
\midrule
W + R & 0.881 $\pm$ 0.033 & 0.972 $\pm$ 0.023 & 0.903 $\pm$ 0.033 & 0.919 $\pm$ 0.023 \\
W + L & 0.879 $\pm$ 0.013 & 0.963 $\pm$ 0.025 & 0.904 $\pm$ 0.018 & 0.915 $\pm$ 0.012 \\
R + L & 0.851 $\pm$ 0.050 & 0.968 $\pm$ 0.015 & 0.884 $\pm$ 0.053 & 0.901 $\pm$ 0.036 \\
\midrule
W+R+L, avg. & 0.880 $\pm$ 0.037 & 0.976 $\pm$ 0.016 & 0.892 $\pm$ 0.029 & 0.916 $\pm$ 0.024 \\
\textbf{W+R+L, stack} & \textbf{0.895 $\pm$ 0.031} & \textbf{0.978 $\pm$ 0.019} & \textbf{0.918 $\pm$ 0.016} & \textbf{0.930 $\pm$ 0.021} \\
\bottomrule
\end{tabular}%
}
\end{table}

\noindent\textbf{Insight.}
\textit{No single view fully captures the architecture fingerprint. View W captures multi-scale activity concentration, View R captures round cadence and regularity, and View L captures round-to-round evolution.}

\subsubsection{Fusion Strategy}
\label{sec:ablation_fusion}

Table~\ref{tab:view_ablation_detailed} also compares average fusion and learning based fusion (stacking). As our dataset is balanced so, the average and weighted-average fusion both achieve a macro F1 of 0.916, hence we show average fusion result. With compare to average fusion, stacking improves macro F1 to 0.930 and yields the best overall result.

\noindent\textbf{Insight.}
\textit{The similar performance of average and weighted fusion suggests that the three views contribute comparably. Stacking performs best because it learns when to trust each view, rather than assigning some fixed weight to every window.}

\subsubsection{Observation Window Analysis}
\label{sec:ablation_window}

We evaluate how the sniffing window affects \textsc{Flint}'s performance by varying
$w$ from 60\,s to 600\,s. Short windows may not contain enough FL round-cycle
structure, while longer windows increase the adversary's observation cost. In real scenario, attack window is not fixed, attacker can come and join any time, hence the impact of sniffing window represent the attack impacts with compare to sniffing time.

\begin{figure}[t]
  \centering
  \includegraphics[width=0.8\columnwidth]{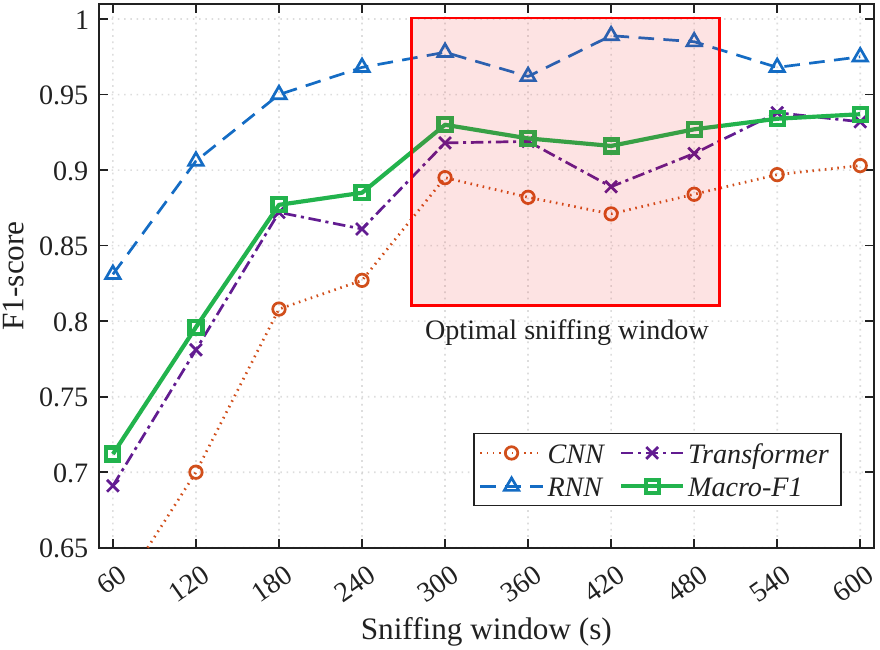}
  \caption{Effect of sniffing window size on \textsc{Flint}'s fingerprinting performance. F1-score improves rapidly from 60\,s to 300\,s as longer windows capture more FL round-cycle structure, and then largely saturates. We use $w=300$\,s as the default operating point because it provides a strong tradeoff between attack accuracy and observation cost.}
  \label{fig:window_f1}
\end{figure}

\noindent\textbf{Insight.}
\textit{\textsc{Flint} does not require observing the full training session. A few minutes of passive PDCCH sniffing are sufficient to recover a reliable architecture fingerprint, while very short observations fail to capture enough FL round-cycle structure.}

\subsection{Exploratory Open-World Rejection Analysis}
\label{sec:openworld}

In practical deployments, the adversary may observe traffic that does not belong to any of the architecture families represented during classifier training. We therefore examine whether the three-class FLINT classifier can reject unknown observations using a confidence threshold. This analysis is intended to evaluate the limitations of standard softmax-based rejection of open-world evaluation.

\paragraph{Rejection mechanism.}

We evaluate confidence-threshold-based rejection using the fused three-class softmax probabilities. An observation window is rejected as \textit{Others} when its maximum class probability is below a threshold $\eta$; otherwise, it is assigned to the architecture family with the highest probability. The threshold $\eta$ is calibrated using held-out validation windows from the known architecture families to retain approximately 90\% of known validation observations. Thus, \textit{Others} is a rejection outcome rather than a separately trained fourth class.


\paragraph{Metrics.}
We report three metrics. \textit{Known Macro F1} is the macro-averaged F1-score over CNN, RNN, and Transformer observations accepted as known. \textit{Unknown Recall} is the fraction of true \textit{Others} observations correctly rejected. The \textit{False Acceptance Rate} (FAR) is the fraction of true \textit{Others} observations incorrectly accepted as one of the known architecture families.


\begin{table}[t]
\centering
\caption{Threshold-based unknown rejection at w=300 s. Known Macro F1 measures classification of known architecture families.}
\label{tab:ow-mechanism}
\begin{tabular}{lccc}
\toprule
Mechanism & Unknown Recall & FAR & Known Macro F1 \\
\midrule
Softmax threshold & 0.111 & 0.889 & \textbf{0.909} \\
\bottomrule
\end{tabular}
\end{table}


\paragraph{Results.}
Table~\ref{tab:ow-mechanism} reports the performance of confidence-threshold-based rejection at $w=300$~s. The thresholded softmax maintains a Known Macro F1-score of 0.909 for accepted known observations. However, it correctly rejects only 11.1\% of \textit{Others} observations and produces an FAR of 0.889. These results indicate that the three-class softmax classifier remains overconfident when presented with observations outside the known architecture families. Consequently, confidence thresholding alone is insufficient for reliable unknown-workload detection.


\noindent\textbf{Insight.}
\textit{Strong classification performance among known architecture families does not automatically provide reliable rejection of unknown observations. The low Unknown Recall and high FAR show that standard softmax confidence is not a dependable indicator of whether a PHY-layer trace belongs to a known architecture family. More specialized open-set recognition methods are therefore required for reliable open-world fingerprinting.}
\vspace{-0.05in}

\subsection{Downstream Attacks Based on Fingerprinting}
\label{sec:downstream}

This subsection establishes the downstream stage of the attack, in which the adversary converts model architecture into surgical interference against the targeted FL client(s). The upstream fingerprinting stage has already revealed to the adversary which architecture the clients run, so the adversary no longer treats the radio link as an opaque pipe and instead jams the uplinks of selected clients at the PHY layer during FL training.

\subsubsection{Attack on Synchronous Federated Learning}
\label{sec:attack-sync}

\begin{figure}[t]
  \centering
  \begin{subfigure}[t]{0.49\linewidth}
    \centering
    \includegraphics[width=\linewidth]{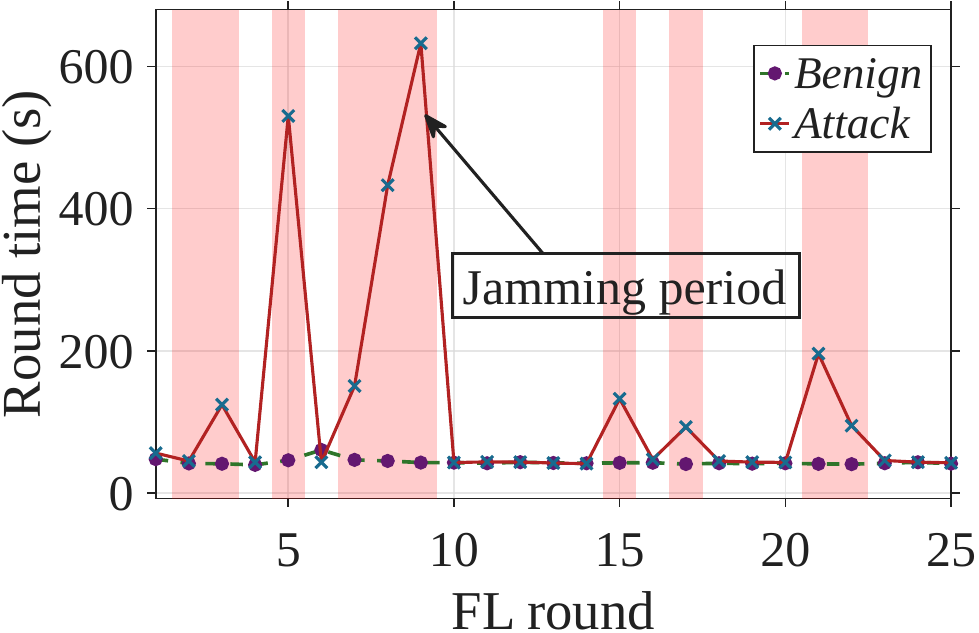}
    \caption{CNN}
    \label{fig:sync-cnn}
  \end{subfigure}
  \hfill
  \begin{subfigure}[t]{0.49\linewidth}
    \centering
    \includegraphics[width=\linewidth]{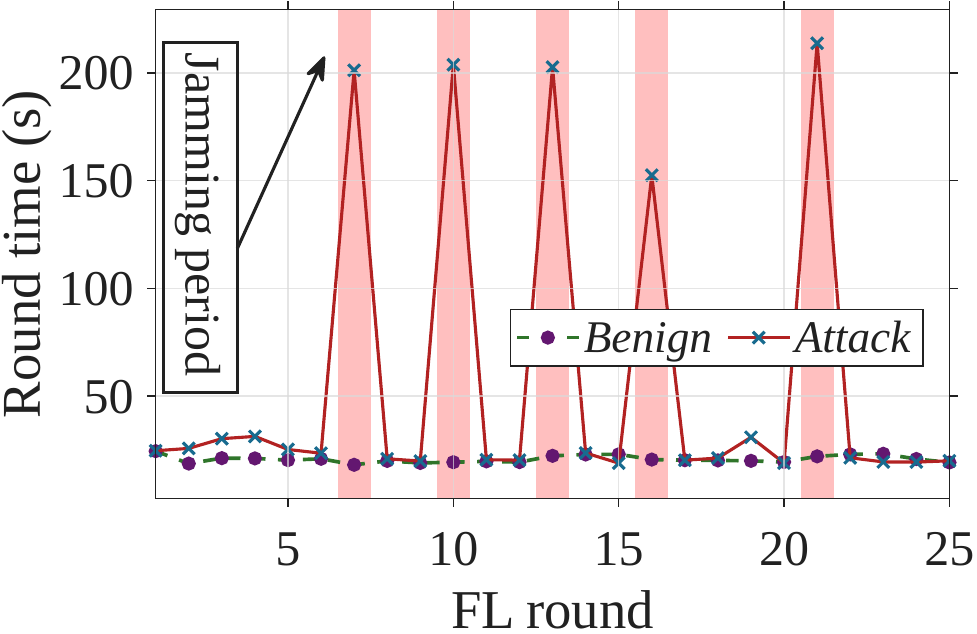}
    \caption{RNN}
    \label{fig:sync-rnn}
  \end{subfigure}
  \caption{Targeted jamming under \emph{synchronous} FL, showing
  round completion time for (a)~CNN and
  (b)~RNN client.}
  \label{fig:sync}
\end{figure}
We first examine synchronous aggregation, where the attack converts a single jammed client into a stall of the entire round, and both architectures suffer. The mechanism follows directly from the synchronization barrier in FedAvg. Because the aggregator cannot produce the next global model until every selected client has returned its update, jamming the targeted client stalls communication between clients and the server. For the convolutional client in Fig.~\ref{fig:sync-cnn}, jammed rounds complete in a mean of $243\,\text{s}$ against a clean-round mean of $44\,\text{s}$, an inflation of roughly $5.5\times$ per jammed round, and the cumulative training time rises by $+152\%$ over the benign baseline. 

Moreover, the recurrent client incurs a higher relative cost. Fig.~\ref{fig:sync-rnn} reports jammed rounds at a mean of $195\,\text{s}$ against a clean-round mean of only $26\,\text{s}$, an inflation of nearly $7.6\times$, with the total training time rising by $+157\%$. The mechanism is identical to the CNN case. In both cases, the jamming effect stays confined to the jamming period, because every spike sits inside an orange band and the round time returns to baseline the moment the band ends, and the final accuracy converges to the benign trajectory at $0.894$ against $0.892$ for the CNN client and $0.905$ against $0.918$ for the recurrent one. The attack therefore denies timely convergence rather than accuracy.

\subsubsection{Attack on Asynchronous Federated Learning}
\label{sec:attack-async}

\begin{figure}[t]
  \centering
  \begin{subfigure}[t]{0.49\linewidth}
    \centering
    \includegraphics[width=\linewidth]{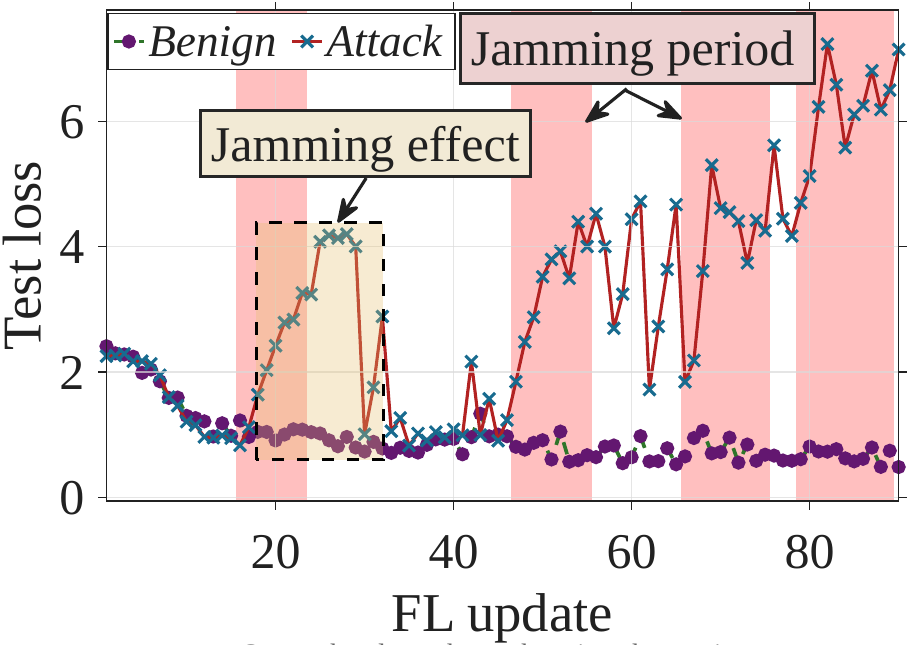}
    \caption{CNN}
    \label{fig:async-cnn}
  \end{subfigure}
  \hfill
  \begin{subfigure}[t]{0.49\linewidth}
    \centering
    \includegraphics[width=\linewidth]{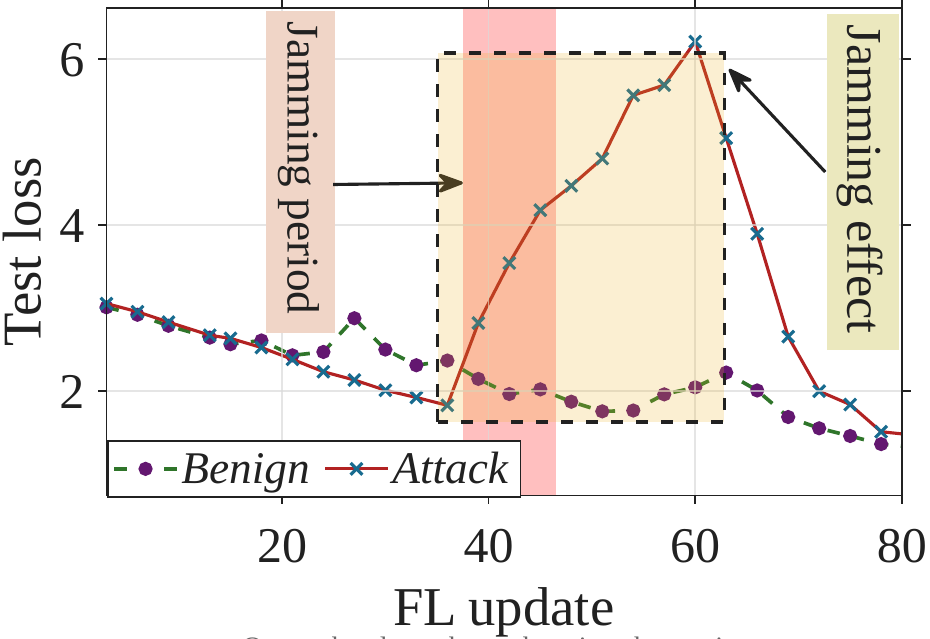}
    \caption{RNN}
    \label{fig:async-rnn}
  \end{subfigure}
  \caption{Targeted jamming under \emph{asynchronous} federated learning, showing
  test loss per asynchronous update for (a)~the convolutional and (b)~the
  recurrent client.} \vspace{-0.25in}
  \label{fig:async}
\end{figure}
We now turn to asynchronous aggregation, where the global model is updated as client updates arrive without waiting. Therefore, jamming no longer stalls progress. Instead, it injects stale and partial contributions into the aggregate. The CNN client in Fig.~\ref{fig:async-cnn} resolves this into two regimes. During the first isolated burst near updates $16$ through $23$, the test loss climbs to roughly $4$ and then falls back toward the benign level within a few updates after the band ends. However, continuous jamming forces the global model to update with fewer client updates, leading to higher loss and a $+6.66$ increase over the baseline. 

The recurrent client in Fig.~\ref{fig:async-rnn} isolates a single short jamming period and exposes a delayed response that outlives it. The jamming period spans only the band near updates $39$ through $45$, yet the test loss continues to climb for roughly $15$ updates after the band ends, peaks near update $60$, and only then decays back toward the benign trajectory, settling a mere $+0.16$ above baseline by the end while the accuracy tracks benign at $0.626$ against $0.628$. The aggregate statistics show the same lag because the jammed updates themselves average only $+1.47$ above benign, while the nominally clean updates average $+0.83$, confirming that much of the damage falls on updates the adversary never touched. 

Therefore, the two cases together support a single structural claim: that synchronous federated learning confines the attack to the time domain within the jamming period while asynchronous federated learning relocates it to the training loss and lets it persist beyond the jamming period, and the magnitude and recoverability within each domain depend jointly on the architecture and the jamming duration. \vspace{-0.05in}

\noindent\textbf{Insight.}
\textit{Fingerprinting makes targeted attack possible. Once the adversary identifies which client is training a particular model family, it can selectively jam that client instead of attacking blindly. The resulting stale updates can then propagate through aggregation, turning model-architecture reconnaissance into a practical downstream attack.}\vspace{-0.05in}

\subsection{Countermeasure}
\label{sec:countermeasure}
\vspace{-0.1 in}
\begin{table}[ht]
\centering
\caption{Effect of countermeasure on \textsc{Flint}.}
\label{tab:defense}
\small
\setlength{\tabcolsep}{3.5pt}
\begin{tabular}{lcccc}
\toprule
Class / Target & Correct & Misclassified & Rejected & Not correctly id. \\
\midrule
CNN / 5 MB & 0.0\% & 66.7\% & 33.3\% & 100.0\% \\
CNN / 10 MB & 85.2\% & 0.0\% & 14.8\% & 14.8\% \\
RNN / 5 MB & 100.0\% & 0.0\% & 0.0\% & 0.0\% \\
RNN / 10 MB & 100.0\% & 0.0\% & 0.0\% & 0.0\% \\
\midrule
Overall & 79.8\% & 7.1\% & 13.1\% & 20.2\% \\
\bottomrule
\end{tabular}
\end{table}

Traffic-analysis defenses commonly use padding, dummy traffic, size normalization, timing obfuscation, burst shaping, and cluster anonymization to reduce side-channel leakage~\cite{dyer2012peek,holland2023detorrent,shen2024real}. We evaluate a lightweight size-obfuscation defense in which each FL update is compressed when possible and then padded to a fixed target size. We test two targets, 5 MB and 10 MB, and measure fingerprinting attack performance.

Table~\ref{tab:defense} shows that size obfuscation provides only limited protection. For CNN with a 5 MB target, none of the defended samples are correctly identified; however, with a 10 MB target, 85.2\% remain correctly identified. RNN is unaffected, with 100.0\% correct identification under both targets. Overall, only 20.2\% of defended samples are either misclassified or rejected. The defense also increases communication overhead from between 14\% - 89\%. These results suggest that effective defenses must hide not only update-size information, but also round-cycle timing and scheduling dynamics exposed through 5G PHY-layer metadata. Existing traffic-analysis countermeasures are largely designed for packet-level fingerprints, where the main observable features are packet size, direction, timing, and burst shape. In contrast, \textsc{Flint} operates on 5G PHY-layer scheduling metadata, where the attacker observes transport block sizes, resource block allocations, transmission direction, and round-cycle dynamics exposed through PDCCH/DCI records. Future defenses should therefore consider 5G-specific obfuscation mechanisms that jointly hide update size, scheduling cadence, uplink/downlink round structure, and resource allocation patterns while keeping the communication overhead practical for federated learning over cellular networks.
\vspace{-0.05in}\vspace{-0.05in}

\section{Conclusion}
\label{sec:conclusion}
This paper presented FLINT, a fully black-box framework for fingerprinting federated learning model architectures from 5G physical-layer side channels. Although 5G encryption protects user-plane packets and AI/ML model update contents, FLINT shows that PDCCH scheduling metadata still leaks architecture-dependent patterns. By reconstructing per-device RNTI grants via a novel RNTI-to-UE mapping, imputing incomplete scheduling observations, and combining multi-resolution energy, round-cadence, and sequential temporal views, FLINT infers whether a client is training a CNN, RNN, or Transformer without requiring participation in training or access to packet-level traffic. Evaluation on a real 5G FL testbed demonstrates that the proposed RNTI mapping accurately recovers device identities and that temporal multi-view fingerprinting substantially outperforms simple PHY-layer statistical baselines. The results further show that open-world rejection and downstream architecture-aware attacks are practical concerns, making this leakage more than a passive privacy issue. Overall, FLINT exposes an overlooked vulnerability at the intersection of 5G control-plane observability and federated learning security, and motivates future defenses that obfuscate or normalize scheduling behavior without degrading FL performance.\vspace{-0.05in}

\bibliographystyle{IEEEtran}
\bibliography{ref_intro,ref_related, ref_others}   
\end{document}